\documentclass[preprint,showpacs,preprintnumbers,amsmath,amssymb,amsfonts]{revtex4}

\usepackage{graphicx}
\usepackage{epsfig}
\usepackage{bm}
\usepackage{verbatim}

\begin{document}

%opening
\title{Long Distance Correlations in Molecular Orientations of Liquid  Water and Shape Dependant  Hydrophobic  Force}
\author{J. M. Pradeep Kanth}
\email{jmpkanth@imsc.res.in}
\author{Satyavani Vemparala}
\email{vani@imsc.res.in}
\author{Ramesh Anishetty} 
\email{ramesha@imsc.res.in}
\affiliation{The Institute of Mathematical Sciences, C.I.T.Campus, Tharamani, Chennai  600113,  India}
\date{\today}

\begin{abstract}
Liquid water, at ambient conditions, has short-range density correlations which are well known in literature. Surprisingly, large scale molecular dynamics simulations reveal an unusually long-distance correlation in `longitudinal' part of dipole-dipole orientational correlations. It is non-vanishing even at $75$ \AA{} and falls-off exponentially with a correlation length of about $24$ \AA{} beyond solvation region. Numerical evidence suggests that the long range nature of dipole-dipole correlation is due to underlying fluctuating network of hydrogen-bonds in the liquid phase. This correlation is shown to give a shape dependant attraction between two hydrophobic surfaces at large distances of separation and the range of this attractive force is in agreement with experiments. In addition it is seen that quadrupolar fluctuations vanish within the first solvation peak ($3$ \AA{})
\end{abstract}

%61.20.Ja 	Computer simulation of liquid structure 
%61.25.Em 	Molecular liquids 
%82.30.Rs 	Hydrogen bonding, hydrophilic effects 
\maketitle

%\section{Introduction}

Water molecule with its hydrogens and lone pairs in tetrahedral arrangement makes hydrogen-bonds with its neighbouring molecules. In the liquid phase the hydrogen-bond pattern undergoes rapid fluctuations at pico-second time scales \cite{StillScience80,StarrPRL99,SayPNAS01,FeckoScience03}, thus resulting in large orientational entropy. It is well known that this special property bestows liquid water with some unique properties, in particular the hydrophobic force of attraction between non-polar solutes. Clever experiments have been performed to measure quantitatively the distance properties of the hydrophobic force between mesoscopic surfaces \cite{IsrPNAS06}. Understanding these distance properties is necessary initial step to develop a proper theory for bulk liquid water. Here we make a preliminary attempt towards the same using molecular dynamics (MD) simulations and general principles of statistical mechanics.

\par
A water molecule can be modelled as a set of five points corresponding to neutral oxygen $O$, two positively polarized hydrogens $H_1$, $H_2$ and two negatively polarized lone-pair sites $L_1$, $L_2$ placed at tetrahedral angles about the oxygen atom. The angles between $\vec{OH}_{1,2}$ and $\vec{OL}_{1,2}$ and the length of each of these vectors can fluctuate. Such a molecule's orientations can be conveniently described with a choice of vectors defined as : 

\begin{equation}
\vec{e}_{1(2)}(r) =  {\frac{\vec{OH_1} + \vec{OH}_2}{|OH_1 + OH_2|}
   - (+)  \frac{\vec{OL}_1 + \vec{OL}_2}{|OL_1  + OL_2|} }  \label{eeqns}
\end{equation}

where `$r$' is the position of oxygen atom in the bulk. The choice of $\vec{e}_1(r)$ and $\vec{e}_2(r)$ is such that they do not depend upon bond lengths of the molecule; they are symmetric with respect to hydrogens and lone-pairs of the molecule. $\hat{e}_1(r)$, $\hat{e}_2(r)$ and $\hat{e}_3  \equiv \hat{e}_1 \times \hat{e}_2$ are the corresponding orthonormal vectors. Here $\hat{e}_1(r)$ is dominantly the direction of dipole field and $\hat{e}_2(r)$ exists only if the water molecule differs from its mean (near-tetrahedral) geometry i.e. it is proportional to the quadrupole moment of the molecule.

The $\hat{e}$-vectors [Eq.~(\ref{eeqns})] form a complete triad with which orientation of any vector ($\vec{OH}$ or $\vec{OL}$ ) can be specified. Consequently dynamics of water can be understood to be an interacting system of the  $\hat{e}$-vector fields. In particular MD simulation of water molecules implicitly gives us the dynamics of these fields. There upon various statistical correlations involving $\hat{e}_1(r)$, $\hat{e}_2(r)$ and $\rho(r) \equiv (\hat{e}_1(r))^2 = (\hat{e}_2(r))^2$ in the liquid phase of water can be formulated as follows :

\begin{subequations}
\label{correqns}
\begin{eqnarray}
< \rho(r_{1}) \rho(r_{2}) >  & = &  g(r) \\*
< \rho(r_{1})  \hat{e}_a(r_{2}) > & = & \frac{\vec{r}}{r}  d_a(r) \\*
\nonumber < e_{a}^{i}(r_{1}) e_{b}^{j}(r_{2}) >  & = &  \frac{1}{2} ( \delta^{ij} - \frac{r^{i} r^{j}}{r^{2}} )  t_{ab}(r) \\  &-&  \frac{1}{2} ( \delta^{ij} - 3 \frac{r^{i} r^{j}}{r^{2}} )  l_{ab}(r) 
\end{eqnarray}
\end{subequations}

where $ \vec{r} = ( \vec{r_1} - \vec{r_2} ) $, $ r = |\vec{r}|$; subscripts $a, b = 1, 2 $ (denote either of $\hat{e}_1, \hat{e}_2 $) and vector indices $i, j = {1, 2, 3 }$ (denote directions in three dimensional space).

The translational and rotational symmetry of the system enables decomposing the tensorial properties of these correlations explicitly and thus analyze the data in terms of simple scalar functions like $g(r)$, $d_a(r)$, $t_{ab}(r)$, $l_{ab}(r)$. The function $g(r)$ is the radial distribution function and it portrays distance-dependant density correlations only (here, of oxygen). The remaining functions capture the correlations among other degrees of freedom of the vector fields.

TIP5P model possesses all orientational degrees of freedom of a water molecule and has improved accuracy in predicting the structural properties of water at ambient conditions. The simulations  of TIP5P water system are performed with GROMACS (version 3.3.1) package \cite{gromacs01} with an integration time step of $2$ fs. The fast-moving bonds $O-H$ are constrained using LINCS algorithm. A large system consisting of $110592$ molecules in a $150$ \AA{} box is equilibrated for $2$ ns in constant pressure (isotropic and $1$ atm) and temperature ($300$ K) $NPT$ ensemble followed by a production run of $2$ ns in a constant volume $NVT$ ensemble. The configurations are saved every $100$ ps for analysis. A cut-off distance of $12$ \AA{} and a pair-list distance of $15$ \AA{} are used to compute all non-bonded interactions and periodic boundary conditions are imposed. Full electrostatic interactions are computed with Particle Mesh Ewald (PME) method  with a tolerance of $10^{-6}$ and updated every two time steps \cite{Frenkel2001, LeachMolMod}.

Density correlation $g(r)$ of TIP5P displays all the well-known solvation peaks; in addition, due to large system size and hence better statistics, few more prominent peaks are observed at about $r = 8.8$ \AA{} and $r = 10.8$ \AA{} \cite{prlsupp}. The orientations of (dipolar) field $\hat{e}_1$ are analyzed by the correlations  $< {e_1}^i(0) {e_1}^j(r) > $ [Eq.~(\ref{correqns})] where $i,j$ refer to components of $\hat{e}_1$ vector. This is conveniently decomposed into two parts : transverse trace part $t_{11}(r) = < \hat{e}_1(0) \cdot \hat{e}_1(r)>$ which measures the dipoles' alignment with respect to each other and thus solely contributes to Kirkwood dielectric function \cite{Kirkwood39,ParriPRL99,ManuPRL07}; and longitudinal traceless part $l_{11}(r) = < \hat{e}_1(0) \cdot \hat{r} \ \hat{e}_1(r) \cdot \hat{r} > $ which is a measure of alignment of the vectors with respect to radial vector separating them.

\begin{figure}
\epsfig{file=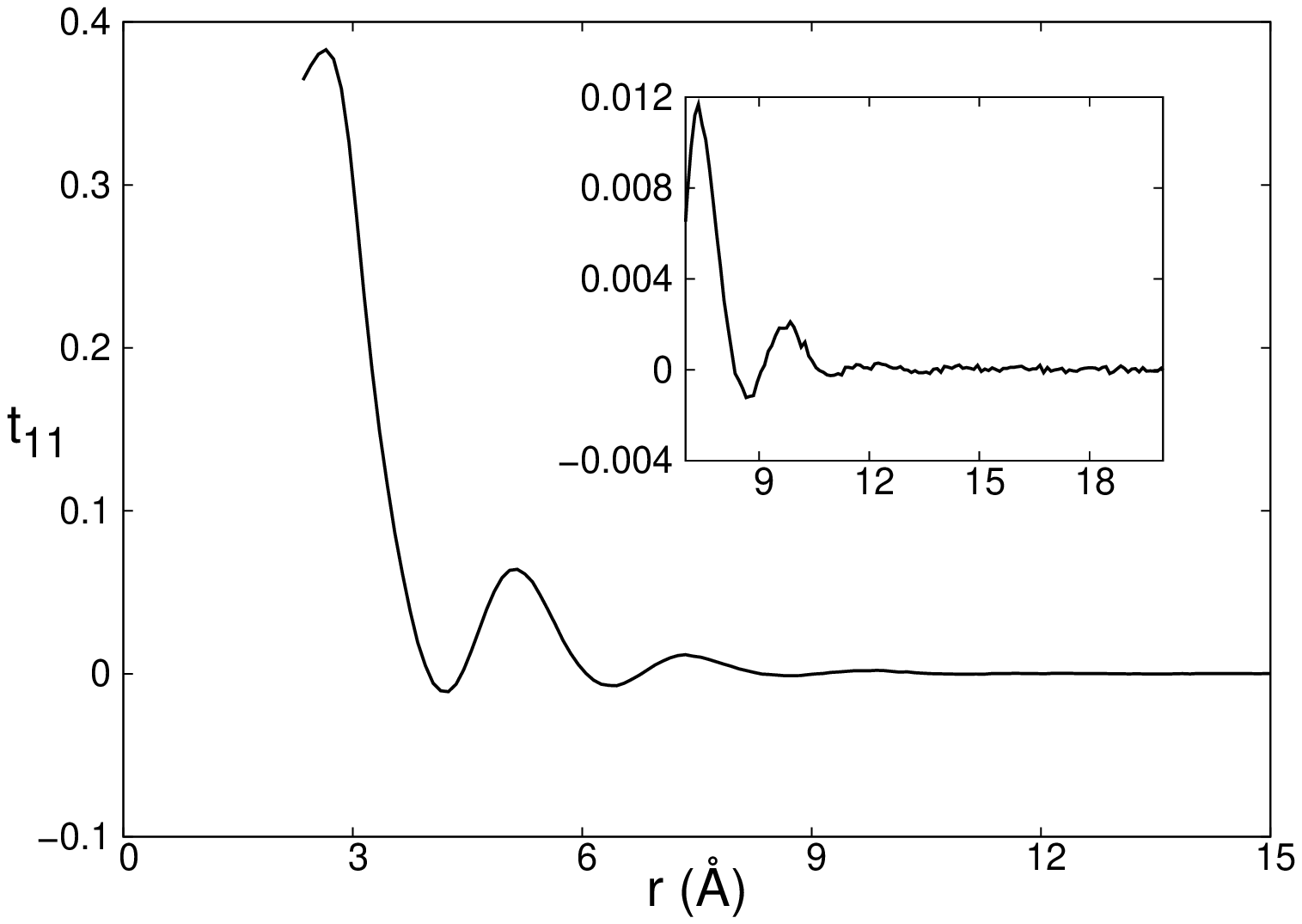}
\caption{\label{t11fig}TIP5P - $t_{11}(r)$. Dipole-dipole transverse trace part showing all the solvation peaks. (\textit{inset}) The correlation vanishes beyond the solvation region of $14$ \AA{}}
\end{figure}

The transverse correlation function $t_{11}(r)$ shows oscillatory solvation structure, but vanishes beyond $14$ \AA{} [Fig.~(\ref{t11fig})] (in compliance with the rotational symmetry in the full system). The function $l_{11}(r)$ is seen to be always positive and furthermore, in the $ 14 - 75 $ \AA{} regime it can be fitted to Ornstein-Zernike (OZ) form as \cite{prlsupp}

\begin{eqnarray}
\nonumber l_{11}(r)  = 0.39(2) \ \frac{e^{\displaystyle {-r}/{5.2(1)}}}{r} +  0.027(1) \ \frac{e^{\displaystyle{-r}/{24(1)}}}{r}   \\  r > 14  \text{\AA{}} \qquad \label{l11eq}
\end{eqnarray}

\begin{figure}
\epsfig{file=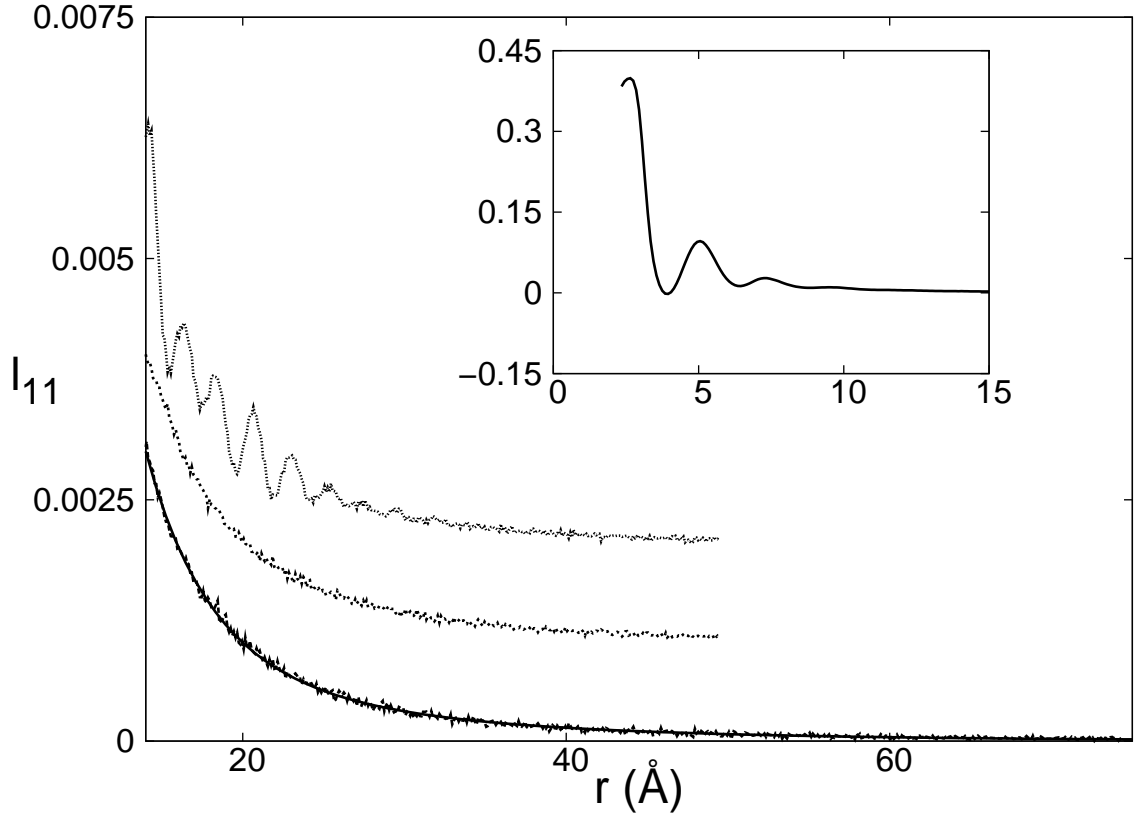}
\caption{\label{l11fig}Exponential decay in longitudinal dipole-dipole correlation $l_{11}(r)$ of liquid water outside the solvation region. (\textit{lower}) TIP5P data and fit function, given by Eq.~(\ref{l11eq}), right on top of each other. (\textit{middle}) TIP3P data. (\textit{upper}) TIP3P with truncated Coulombic interactions. For clarity, the middle and upper plots are shifted up by $0.001$ and $0.002$ units respectively. (\textit{inset})  $l_{11}(r)$ inside the solvation region.}
\end{figure}

$l_{11}(r)$ shows longest correlation length of $24$ \AA{}. Furthermore it exhibits solvation peaks upto $14$ \AA{} [Fig.~(\ref{l11fig})]. In our simulation data upto $75$ \AA{} the statistical sampling errors dramatically reduce as we go to large distances (as expected) \cite{prlsupp}.

The dipole-oxygen correlation  $d_1(r)  =  <\hat{e}_1(0) \cdot \hat{r} \rho(r)>$ also exhibits solvation structure and vanishes beyond $14$ \AA{}. It is also found that correlations involving $\hat{e}_2$, $\hat{e}_3$ all vanish upto statistical errors beyond the first solvation peak \cite{prlsupp}. Therefore $\hat{e}_2$, the quadrupole moment of water, fluctuates locally and randomly without any non-local correlations.

TIP3P model \cite{JorgJCM83}, by design, has $ \hat{e}_1 $ degree of freedom only i.e. each water molecule's orientation can be completely described by $  \hat{e}_1 $ field alone. The simulations on TIP3P water system are performed using NAMD (version 2.6) \cite{Kale98namd}. Here, $33105$ water molecules are simulated in a cubical box of size $100$ \AA{} and the procedures employed for collecting equilibriated configurations are same as those described in case of TIP5P. The constrained model is implemented using SETTLE algorithm. Analysis in this case too shows that $t_{11}(r)$ vanishes beyond solvation region, whereas $l_{11}(r)$ follows the same asymptotic behaviour as described by Eq.~(\ref{l11eq}).

A water molecule in liquid phase is predominantly influenced by hydrogen-bonding (short-range interaction) and further, it has a net dipole moment, which interacts through long-range Coloumbic forces. We would like to ascertain if the long distance behaviour of $l_{11}(r)$ is due to the short-range hydrogen-bond interactions or the long-range Coloumbic interactions \cite{HansenSimpleLiquids,TavJCP04}. To test this possibility, the Coulombic interactions are smoothly truncated at $12$ \AA{} in TIP3P model, thus retaining an effective short-range interaction alone which imitates the effect of hydrogen-bonding. We find that $l_{11}(r)$ remains essentially unchanged in the regions of first few solvation shells and $r \ > \ 30$ \AA{}. The intermediate region exhibits over-structuring effects upto $30$ \AA{} \cite{AllenLiquids87,KolafaJMolLiq00}.

The above three cases are in agreement with Eq.~(\ref{l11eq}) asymptotically. These observations suggest that (i) water in liquid phase has fluctuations only in dipole degree of freedom; in contrast the quadrupole has no effect beyond the first solvation peak, (ii) these dipole fluctuations in liquid water are influenced by local environment of respective molecule, through hydrogen-bonding, significantly more compared to long-range electrostatic interactions, (iii) furthermore, the dipole fluctuations exhibit long distance correlations. Simulations show that OZ like behaviour is exhibited only by longitudinal dipole-dipole correlation and not transverse component or oxygen-oxygen density correlation. Recently there has been an experimental observation which reported correlation length of the order of a nanometer in liquid water \cite{HuangPNAS09}; in line with the shorter correlation length in $l_{11}(r)$. Below we suggest that the longer correlation length of $24$ \AA{} is indirectly observed in SFA experiments as force between hydrophobic plates.

\textbf{Hydrophobic effect} : The first notable mechanism postulated to describe the origin of hydrophobic force came from solvation studies of Frank and Evans in the name of ``iceberg'' model \cite{Frank45} and later, the same effect has been elucidated by Kauzmann on its possible biological implications \cite{Kauz59}. This phenomenon, in its various manifestations, has been extensively discussed in recent literature \cite{BallChemRev08,Chaplin06,DillAnnRev05,ChandlerNat05}. Experimentally the force of attraction between two nominally hydrophobic macroscopic surfaces has been measured and it was found that in the range $10$ \AA{} to $100$ \AA{} the force falls-off exponentially with a correlation length of $12$ \AA{} \cite{IsrNature82}. Later there have been several such studies using Surface Force Apparatus (SFA) with surfaces prepared and characterized using wide range of techniques \cite{Claess01,IsrPNAS06}. Yet there were very few theoretical developments (Lum \textit{et al} \cite{LumJPC99} and ref's therein, \cite{DespaPRL04,Besseling97}) to explain qualitatively different force profiles observed in experiments. We address here monotonic nature of the force as observed in several experiments \cite{IsrNature82,Claess01,IsrPNAS06}.

Hydrophobic surfaces cannot form hydrogen-bonds with the surrounding water, consequently water molecules rearrange themselves such that they form a sheet of hydrogen-bond network on the surface. Their interactions are such that the directions of lone pairs and hydrogen atoms are perpendicular to the surface normal of the hydrophobe. Owing to the approximate tetrahedral conformation, water molecules cannot have a unique configuration satisfying the above criterion \cite{RossNature98}. Consequently they explore other possible orientations as well by fluctuating at the pico-second time scales \cite{RezPRL07,PoynorPRL06}. These network fluctuations contribute significantly to the free energy of solvation of the hydrophobe. In presence of two such hydrophobes, as noted earlier, the range of the force acting between them is large and due to limited computational resources it is not possible to directly simulate and observe this effect numerically. Alternatively a quantitative theoretical estimate is being considered below.

Interaction between hydrophobic surface and solvent water can be written in terms of $\hat{n}(r)$, the local unit normal vector to the hardcore van der Waals surface of the hydrophobe and $\hat{e}_1(r')$, the dipole of water molecule near the surface, where $r' = r+ \delta r$; $\delta r$ is typical length of hydrogen arm of water molecule (about $1$ \AA{}). A simple local interaction term can be taken as $( \hat{n}(r) \cdot \hat{e}_1(r') )^2 $ implying that the water dipoles orient orthogonal to the surface normal as seen in simulations \cite{Chau01,JedJPhys04,LevittPNAS05} (importantly, no linear term in $\hat{n} \cdot \hat{e}_1$, for that means a preferential orientation of the water dipole inward/outward to the surface).

\begin{figure}
\epsfig{file=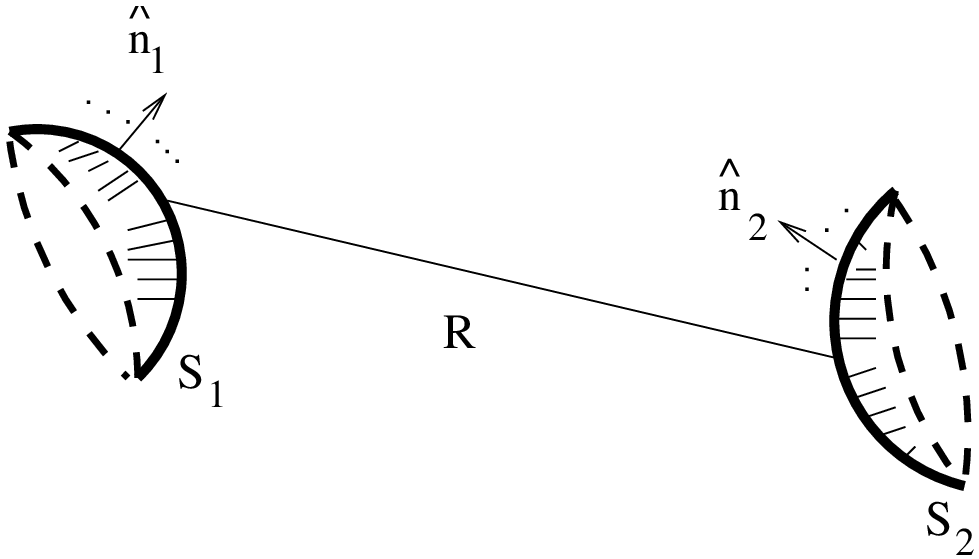}
\caption{\label{surfacefig}$S_1, S_2$ are hydrophobic surfaces with their local normal vectors $\hat{n}_{1}$, $\hat{n}_{2}$. '$R$' is the minimum distance between the two surfaces}
\end{figure}

The change in free energy due to purely hydrophobic interaction between two small surfaces $S_1$ and $S_2$ [Fig.~(\ref{surfacefig})] in water can be estimated by,

%\begin{widetext}
\begin{eqnarray}
\nonumber \Delta H &=& \frac{\gamma_{1}}{2}   \int_{S_1} d\hat{n}_1  (\hat{n}_{1}(r_1) \cdot \hat{e}_1(r'_{1}))^2  \\ \nonumber 
&+&  \frac{\gamma_{2}}{2} \int_{S_2} d\hat{n}_2  (\hat{n}_{2}(r_2) \cdot \hat{e}_1(r'_{2}))^2 
\\
e^{-{\Delta G}/{kT}}  &=&  < e^{-{\Delta H}/{kT}} > \label{energyeq} 
\end{eqnarray}
%\end{widetext}

where $\gamma$ is a measure of strength of interaction between hydrophobic solute and water which can depend upon temperature, density and other parameters defining the thermodynamic system. The brackets $<...>$ refer to statistical averaging with respect to pure water system and integration is over area of each surface. As illustrated in Fig.~(\ref{surfacefig}), $S_{1}$, $S_{2}$ refer to two arbitrary hydrophobic surfaces and $\vec{R}$ is a vector along minimum distance of separation between them.

When the distance $R$ ($= |\vec{R}|$) is large compared to radius of curvature of each surface and the surface areas sufficiently small, the statistical averaging can be done by cumulant expansion. The leading term of the force $F(R) = -{\partial \Delta G}/{\partial R}$ is given by the following equation \cite{prlsupp},

%\begin{widetext}
 \begin{eqnarray}
\nonumber F(R) &\simeq& \frac{\gamma_1 \gamma_2}{2kT} \frac{\partial}{\partial R} [ \int_{S_1} \int_{S_2} d\hat{n}_1  d\hat{n}_2 \\
\nonumber &&  <[\hat{n}_1(r_1) \cdot \hat{e}_1(r'_{1}) \ \hat{n}_2(r_2) \cdot \hat{e}_1(r'_{2})]^2> ] 
\\
 &=&  \frac{\gamma_1 \gamma_2}{2kT}  A_1  A_2 \ \frac{\partial}{\partial R}  Tr[ \mathbf{\Sigma_{S_1}  E}(R)  \mathbf{\Sigma_{S_2}  E}(R) ]  \label{forceeq}
\end{eqnarray}
%\end{widetext} 

where $A_1$, $A_2$ are areas of the surfaces and the matrices $\mathbf{E(\mathnormal{R})}$, $\mathbf{\Sigma_S}$  are given by, 

\begin{eqnarray*}
E^{ij}(R) & \equiv & < {e_1}^i(r'_{1}) \ {e_1}^j(r'_{2}) > \\ 
&\simeq& - \frac{1}{2} ( \delta^{ij} - 3 \frac{R^i R^j}{R^2} ) {l_{11}}(R) \qquad \text{for large R} \\
\Sigma_{S}^{ij} &\equiv& \frac{1}{A} \int_S d\hat{n} \ {n}^i {n}^j
\end{eqnarray*}
 
For a segment of spherical surface (such as a hardcore van der Waals surface) subtending a cone angle $\theta$ at its centre and $N^i = {M^i}/{|M|}$ where $M^i = \frac{1}{A} \int_{S} d\hat{n} \ n^i $, $ \Sigma_{S}^{ij} = \frac{1}{3} \delta^{ij} - \frac{1}{6} ( \delta^{ij} - 3 N^i N^j ) cos\theta  (1 + cos\theta )$. For a hemi-sphere, $\theta = \frac{\pi}{2}$, $ \Sigma^{ij} = \frac{1}{3} \delta^{ij} $. For a plane, $\theta = 0 $, $ \Sigma^{ij} = N^i N^j $ \cite{prlsupp}. 

The above result on hydrophobic force is true very generally. As discussed in earlier paragraphs, the leading order  $(\hat{n} \cdot \hat{e}_1)^2$ is taken to be the interaction energy term for simplicity. By including the non-leading terms in the interaction energy function [Eq.~(\ref{energyeq})] and doing the cumulant expansion, it can be shown that the force term [Eq.~(\ref{forceeq})] for large $R$ remains unchanged, thus establishing the generality of the result. 

These considerations are valid for distances beyond the solvation region of a typical water molecule. The cumulant expansion allowed decomposing the force equation as a simple convolution of surface-dependant part  and water-dependant part. Eq.~(\ref{forceeq}) enables us to conclude that range of the force between hydrophobic surfaces at large distances is always attractive governed by $ l_{11}^2(R) \simeq e^{-R/12} $ for large $R$. Therefore the hydrophobic force falls off exponentially with a largest correlation length of about $12$ \AA{}, in addition to several other shorter range exponents as well. 

\begin{equation}
F(R)  \sim  -e^{-R/12} \qquad \text{for large R} \label{forceform}
\end{equation}

The strength of attraction is proportional to area $A$ and shape of each surface given by the tensor $\mathbf{\Sigma}$, the second moment of surface normal. The final trace operation over the matrices $\mathbf{E(\text{R})}$ and $\mathbf{\Sigma_S}$ [as in Eq.~(\ref{forceeq})] implies that the hydrophobic attraction is not just a purely distance dependant interaction like van der Waals'. Indeed the orientation of the surface shapes relative to each other can modify this force significantly. As an example if two small planar hydrophobic surfaces are mutually perpendicular and are sufficiently far apart, there should be no force between them as opposed to when they face each other.

\pagebreak

\begin{thebibliography}{36}

\bibitem{StillScience80}
F.~H. Stillinger, Science {\bf 209},  451  (1980).

\bibitem{StarrPRL99}
F.~W. Starr, J.~K. Nielsen, and H.~E. Stanley, Phys. Rev. Lett. {\bf 82},  2294
   (1999).

\bibitem{SayPNAS01}
F.~N. Keutsch and R.~J. Saykally, Proc. Natl. Acad. Sci. U. S. A. {\bf 98},
  10533  (2001).

\bibitem{FeckoScience03}
C.~J. Fecko {\it et~al.}, Science {\bf 301},  1698  (2003).

\bibitem{IsrPNAS06}
E.~E. Meyer, K.~J. Rosenberg, and J.~N. Israelachvili, Proc. Natl. Acad. Sci.
  U. S. A. {\bf 103},  15739  (2006).

\bibitem{gromacs01}
E. Lindahl, B. Hess, and D. van~der Spoel, J. Mol. Mod. {\bf 7},  306  (2001).

\bibitem{Frenkel2001}
D. Frenkel and B. Smit, {\em Understanding Molecular Simulation (Computational
  Science Series, Vol 1)} (Academic Press, {U.}{S.}{A.}, 2001).

\bibitem{LeachMolMod}
A. Leach, {\em Molecular Modelling: Principles and Applications (2nd Edition)}
  ({Prentice Hall}, {U.}{S.}{A.}, 2001).

\bibitem{prlsupp}
See supporting information at www.imsc.res.in/$\sim$maruthi/research.html

\bibitem{Kirkwood39}
J.~G. Kirkwood, J. Chem. Phys. {\bf 7},  911  (1939).

\bibitem{ParriPRL99}
P.~L. Silvestrelli and M. Parrinello, Phys. Rev. Lett. {\bf 82},  3308  (1999).

\bibitem{ManuPRL07}
M. Sharma, R. Resta, and R. Car, Phys. Rev. Lett. {\bf 98},  247401  (2007).

\bibitem{JorgJCM83}
W.~L. Jorgensen {\it et~al.}, J. Chem. Phys. {\bf 79},  926  (1983).

\bibitem{Kale98namd}
L. Kale {\it et~al.}, J. Comput. Phys. {\bf 151},  283  (1999).

\bibitem{HansenSimpleLiquids}
J.-P. Hansen and I.~R. McDonald, {\em Theory of Simple Liquids, Third Edition}
  ({Academic Press}, {U.}{K.}, 2006).

\bibitem{TavJCP04}
G. Mathias and P. Tavan, J. Chem. Phys. {\bf 120},  4393  (2004).

\bibitem{AllenLiquids87}
M.~P. Allen and D.~J. Tildesley, {\em Computer Simulation of Liquids}, {\em
  Oxford Science Publications}, 1 ed. (Clarendon Press, Oxford, 1987).

\bibitem{KolafaJMolLiq00}
J. Kolafa and I. Nezbeda, Mol Phys {\bf 98},  1505  (2000).

\bibitem{HuangPNAS09}
C. Huang {\it et~al.}, Proc. Natl. Acad. Sci. U. S. A. {\bf Epub before print
  \textnormal{doi: 10.1073/pnas.0904743106 (2009)}},    (The authors attribute
  the correlation length to density fluctuations. This aspect needs to be
  investigated in more detail.).

\bibitem{Frank45}
H.~S. Frank and M.~W. Evans, J. Chem. Phys. {\bf 13},  507  (1945).

\bibitem{Kauz59}
W. Kauzmann, Adv. Protein Chem. {\bf 14},  1  (1959).

\bibitem{BallChemRev08}
P. Ball, Chem. Rev. {\bf 108},  74  (2008).

\bibitem{Chaplin06}
M. Chaplin, Nat. Rev. Mol. Cell Biol. {\bf 7},  861  (2006).

\bibitem{DillAnnRev05}
K.~A. Dill, T.~M. Truskett, V. Vlachy, and B. Hribar-Lee, Annu. Rev. Biophys.
  Biomol. Struct. {\bf 34},  173  (2005).

\bibitem{ChandlerNat05}
D. Chandler, Nature {\bf 437},  640  (2005).

\bibitem{IsrNature82}
J.~N. Israelachvili and R. Pashley, Nature {\bf 300},  341  (1982).

\bibitem{Claess01}
H.~K. Christenson and P.~M. Claesson, Adv. Colloid Interface Sci. {\bf 91},
  391  (2001).

\bibitem{LumJPC99}
K. Lum, D. Chandler, and J.~D. Weeks, J. Phys. Chem. B {\bf 103},  4570
  (1999).

\bibitem{DespaPRL04}
F. Despa, A. Fernandez, and R.~S. Berry, Phys. Rev. Lett. {\bf 93},  228104
  (2004).

\bibitem{Besseling97}
N.~A.~M. Besseling, Langmuir {\bf 13},  2113  (1997).

\bibitem{RossNature98}
Y.~K. Cheng and P.~J. Rossky, Nature {\bf 392},  696  (1998).

\bibitem{RezPRL07}
Y.~L.~A. Rezus and H.~J. Bakker, Phys. Rev. Lett. {\bf 99},  148301  (2007).

\bibitem{PoynorPRL06}
A. Poynor {\it et~al.}, Phys. Rev. Lett. {\bf 97},  266101  (2006).

\bibitem{Chau01}
P.~L. Chau, Mol. Phys. {\bf 99},  1289  (2001).

\bibitem{JedJPhys04}
P. Jedlovszky, J. Phys.: Condens. Matter {\bf 16},  5389  (2004).

\bibitem{LevittPNAS05}
T.~M. Raschke and M. Levitt, Proc. Natl. Acad. Sci. U. S. A. {\bf 102},  6777
  (2005).

\end{thebibliography}
\end{document}